\documentclass[conference]{IEEEtran}

\usepackage[utf8]{inputenc}
\usepackage[T1]{fontenc}
\usepackage{cite}
\usepackage{graphicx}
\usepackage{booktabs}
\usepackage{array}
\usepackage{multirow}
\usepackage[table]{xcolor}
\usepackage{tabularx}
\usepackage{tikz}
\usetikzlibrary{arrows.meta,positioning}
\usepackage{pgfplots}
\pgfplotsset{compat=1.17}
\usepackage{url}
\usepackage[hidelinks]{hyperref}

\newcolumntype{L}[1]{>{\raggedright\arraybackslash}p{#1}}
\newcolumntype{C}[1]{>{\centering\arraybackslash}p{#1}}
\graphicspath{{figures/}}
\newcolumntype{Y}{>{\raggedright\arraybackslash}X}
\newcommand{\utwonet}{\mbox{U$^2$-Net}}  
\definecolor{segc}{RGB}{58,110,186}
\definecolor{matc}{RGB}{226,128,40}
\definecolor{pipc}{RGB}{62,150,90}
\definecolor{grp}{gray}{0.92}

\newcommand{\tile}[3]{%
  \begin{tabular}[t]{@{}c@{}}%
    {\setlength{\fboxsep}{0pt}\setlength{\fboxrule}{0.4pt}%
     \fcolorbox{gray!55}{white}{\includegraphics[width=#1]{#2}}}\\[1pt]%
    {\scriptsize #3}%
  \end{tabular}}

\begin{document}

\title{Selecting Person Segmentation Models for\\ Real-Time Video Call Backgrounds on the CPU}

\author{\IEEEauthorblockN{Kawshik Kumar Paul}
\IEEEauthorblockA{Independent Researcher\\
B.Sc.(Engg.) in CSE, BUET\\
kawshikbuet17@gmail.com}}

\maketitle

\begin{abstract}
Background blur and background replacement are now normal parts of a video call. Many open models and open source pipelines can do this, but it is hard to know which one to pick. Speed, edge quality and license all matter, and they are often reported on different hardware, or not reported at all. In this work we ran 20 setups on the same laptop CPU, with the same 640x480 frames and the same output view. The setups are 15 open models and 5 complete open source pipelines. We measured frames per second on a test clip and on a live webcam, compared the masks on a public domain portrait and on three video call photos with a raised hand, one of them a close front camera view, and recorded the license of every model. The fastest setups reach more than 40 frames per second on a laptop CPU, while the slowest, a high detail model for still images, needs several seconds per frame. A few setups give clean edges and still run in real time. We also found that the runtime library and the steps around the model can change the speed by several times. Based on these results we give a guide that tells the reader which model to choose for a given need, such as a weak laptop, the best hair edges, a closed source product, or people behind the speaker.
\end{abstract}


\section{Introduction}
\label{sec:intro}

\begin{figure}[t]
  \centering
  \tile{0.235\columnwidth}{input.jpg}{(a) Input}\hfill
  \tile{0.235\columnwidth}{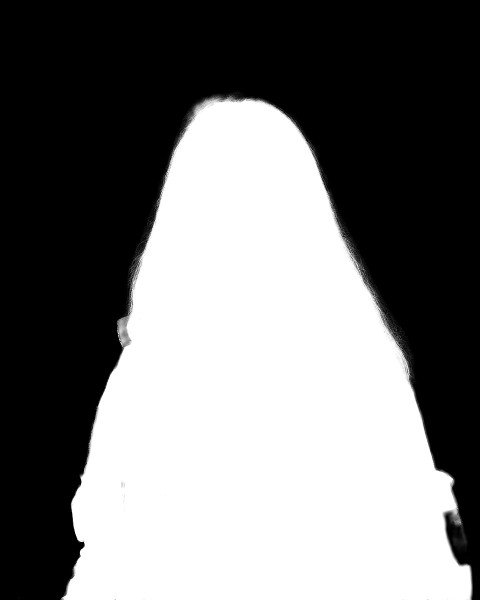}{(b) Mask}\hfill
  \tile{0.235\columnwidth}{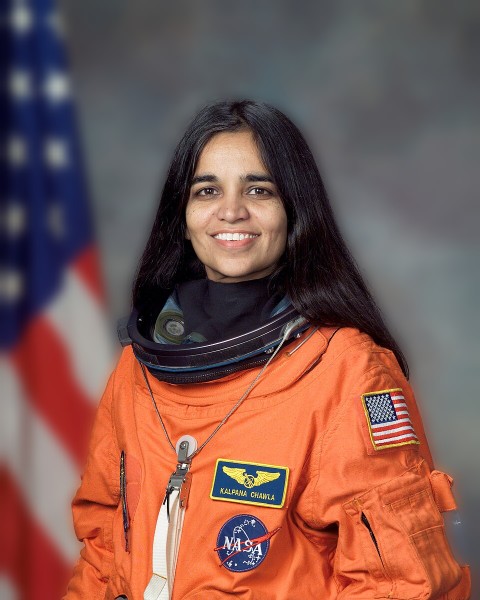}{(c) Blur}\hfill
  \tile{0.235\columnwidth}{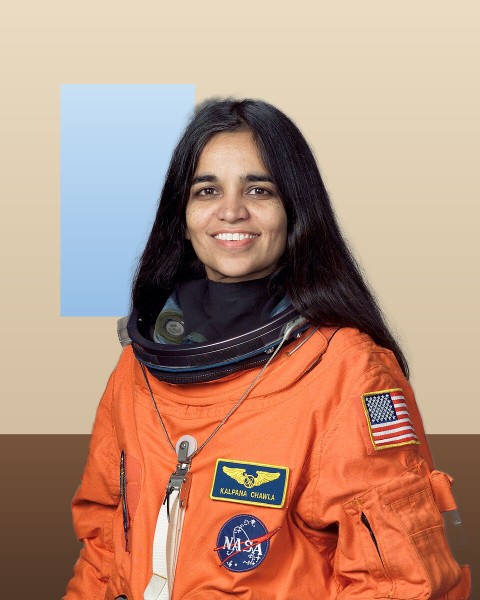}{(d) Replace}
  \caption{The task. A person mask (b) is predicted for the input frame (a) and used to blur the background (c) or to replace it (d). Outputs from RVM. The test image is a public domain NASA portrait of Kalpana Chawla.}
  \label{fig:example}
\end{figure}

In a video call, a virtual background needs a mask that says which pixels belong to the person. With this mask the software can blur the room or place a new image behind the person (\figurename~\ref{fig:example}). The mask has to be ready for every frame, usually 15 to 30 times per second, and it runs next to the video encoder and the call itself. Many computers have no strong GPU, so the mask often has to come from the CPU.

There are many open options for this job. Some are small segmentation networks made for portraits, like PP-HumanSeg~\cite{chu2021pphumanseg}, SINet~\cite{park2019sinet} and the MediaPipe selfie models~\cite{mediapipe}. Some are matting networks that predict soft edges, like MODNet~\cite{ke2020modnet} and Robust Video Matting (RVM)~\cite{lin2021rvm}. Some are general models that also find people, like DeepLabV3~\cite{chen2017deeplabv3} and YOLO~\cite{ultralytics}. There are also complete open source pipelines that add extra steps around a model, for example to make the edge follow the hair or to stop the mask from flickering.

A developer who wants to add a virtual background to a product has to answer simple questions. Which model is fast enough on my users' laptops? Which one keeps the hair and the fingers? Can I use it in a closed source product? What happens when a colleague sits behind me? The answers are spread over many papers and repositories, and their numbers come from different machines, different input sizes and often a GPU.

This paper tries to answer these questions in one place. We ran every setup on the same Windows laptop, on the CPU only, at 640x480. Each setup produces the same four outputs (original, mask, blurred background, replaced background), so the results are easy to compare. We report speed, how the masks look, the license, and the practical problems we met.

Our contributions are the following.
\begin{itemize}
  \item A side by side speed test of 20 setups on one laptop CPU, with both a video file and a live webcam.
  \item A visual comparison of all masks on four photos, with close views of the background leaks, the hair and the fingers of a raised hand.
  \item A license overview, because some popular models cannot be used freely in a closed source product.
  \item A set of practical findings that changed speed or quality a lot, such as the choice of runtime library.
  \item A selection guide that maps common needs to a recommended model (Section~\ref{sec:guide}).
\end{itemize}

Readers who only want a recommendation can go straight to Table~\ref{tab:guide} in Section~\ref{sec:guide}.

\section{Background and Related Work}
\label{sec:background}

\subsection{Segmentation and matting}
A segmentation model gives each pixel a person score between 0 and 1. Some of them give an almost clear yes or no, so the edge is hard and thin hair is often cut away. A matting model predicts the alpha value, which is how much of each pixel belongs to the person. Alpha is soft at the hair and at the edges, so the new background mixes in naturally. Matting models such as MODNet~\cite{ke2020modnet} and RVM~\cite{lin2021rvm} usually look better at the hair but cost more time. RVM also keeps a memory of earlier frames, which makes its mask steady on video.

\subsection{Models made for video calls}
PP-HumanSeg~\cite{chu2021pphumanseg} was built for teleconferencing and comes with a large video conference dataset. SINet~\cite{park2019sinet} targets very small portrait networks. Google has described the small segmentation models behind the background effects in Google Meet~\cite{googlemeetblog}, and it offers selfie segmentation models through MediaPipe~\cite{mediapipe} and ML Kit~\cite{mlkitselfie}. General models such as DeepLabV3~\cite{chen2017deeplabv3} with a MobileNetV3 backbone~\cite{howard2019mobilenetv3} and \utwonet{}~\cite{qin2020u2net} can also produce a person mask, although they were not designed for this task. BiRefNet~\cite{zheng2024birefnet} is a recent high resolution segmentation model with detailed edges.

\subsection{Steps around the model}
\begin{figure}[t]
  \centering
  \begin{tikzpicture}[
      node distance=3.6mm,
      box/.style={draw=#1!70!black, fill=#1!12, rounded corners=3pt, align=center, font=\scriptsize,
                  minimum height=8mm, inner sep=2pt, text width=13mm, line width=0.6pt},
      opt/.style={box=#1, dashed},
      arr/.style={-{Stealth[length=1.8mm]}, line width=0.7pt, gray!70!black}]
    \node[box=gray] (in) {Camera\\frame};
    \node[box=gray, right=of in] (pre) {Resize and\\normalize};
    \node[box=segc, right=of pre, font=\scriptsize\bfseries] (net) {Network};
    \node[opt=pipc, below=6mm of net] (tmp) {Temporal\\smoothing};
    \node[opt=pipc, left=of tmp] (edge) {Edge\\refinement};
    \node[box=matc, left=of edge] (out) {Blur or\\replace};
    \draw[arr] (in) -- (pre);
    \draw[arr] (pre) -- (net);
    \draw[arr] (net) -- (tmp);
    \draw[arr] (tmp) -- (edge);
    \draw[arr] (edge) -- (out);
  \end{tikzpicture}
  \caption{A typical virtual background pipeline. Dashed steps are optional. Complete pipelines often add them around the network.}
  \label{fig:pipeline}
\end{figure}
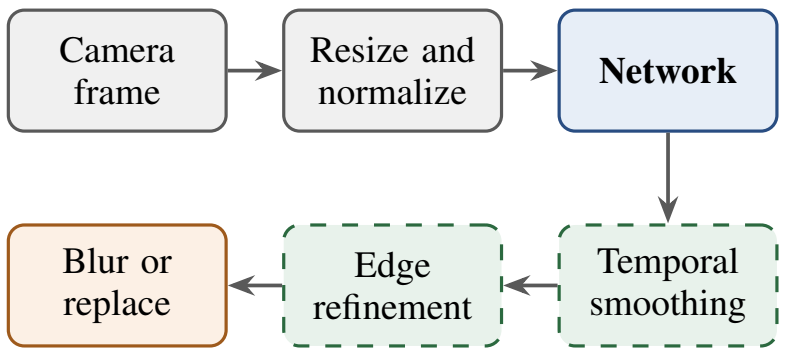

Many open source pipelines add steps before and after the network (\figurename~\ref{fig:pipeline}).
\emph{Temporal smoothing} mixes in the mask of the previous frame, so a still edge does not flicker. Good pipelines follow the motion so that a moving hand does not leave a trail.
\emph{Edge refinement} scales the small mask up to the frame size while looking at the full size picture, so the edge snaps to real edges in the image. A joint bilateral filter is a common way to do this~\cite{kopf2007jbu}.
\emph{Mask clean up} uses thresholds, small blurs or a few pixels of growth to remove specks and holes.
These steps can improve the result a lot, but they also cost time, as Section~\ref{sec:findings} shows.

\section{Benchmarked Methods}
\label{sec:methods}

Table~\ref{tab:setups} lists the 20 setups. Setups 1 to 15 are models that we ran with only the steps their authors describe, for example their own preprocessing. For YOLO we also added two optional steps of our own, one that keeps only the biggest person and one that softens the hard edge. Setups 16 to 20 are complete open source pipelines. The originals of these run in a web browser, on Linux only, or on an NVIDIA GPU, so we rewrote their processing steps in Python. Our versions follow the steps and settings of the originals, but they are not the original code, and their speed is not the speed of the original apps.

\begin{table*}[!t]
  \centering
  \caption{The 20 benchmarked setups. Input is the size of the picture that goes into the network (s means short side, l means long side). Size is the size of the downloaded model files. Runtime versions were PaddlePaddle 3.3.1, MediaPipe 1.0.1, ONNX Runtime 1.23.2~\cite{onnxruntime}, PyTorch 2.14.1, TensorFlow 2.11.0, OpenCV 5.0 (4.11 for setups 10 and 11), ncnn 1.0.20260526, LiteRT 2.2.0 and Ultralytics 8.4.173.}
  \label{tab:setups}
  \footnotesize
  \renewcommand{\arraystretch}{1.15}
  \setlength{\tabcolsep}{4pt}
  \begin{tabularx}{\textwidth}{@{}r Y p{3.3cm} p{2.15cm} p{1.75cm} r p{2.35cm}@{}}
    \toprule
    \textbf{\#} & \textbf{Method} & \textbf{Type} & \textbf{Input} & \textbf{Runtime} & \textbf{Size} & \textbf{License} \\
    \midrule
    \rowcolor{grp}\multicolumn{7}{@{}l}{\textbf{Models}} \\
    1 & PP-HumanSeg v1 Lite~\cite{chu2021pphumanseg,paddleseg} & Segmentation & 398x224 & PaddlePaddle & 2.3 MB & Apache-2.0 \\
    2 & PP-HumanSegV2 Lite~\cite{chu2021pphumanseg,paddleseg} & Segmentation & 256x144 & PaddlePaddle & 5.6 MB & Apache-2.0 \\
    3 & MediaPipe Selfie, landscape~\cite{mediapipe} & Segmentation & 256x144 & MediaPipe & 0.25 MB & Apache-2.0, see model card \\
    4 & MediaPipe Selfie Multiclass~\cite{mediapipe} & Segmentation, 6 classes & 256x256 & MediaPipe & 16.4 MB & Apache-2.0, see model card \\
    5 & MODNet~\cite{ke2020modnet,modnetcode} & Matting & s512 or s256 & ONNX Runtime & 25.9 MB & Apache-2.0 \\
    6 & RVM, MobileNetV3~\cite{lin2021rvm,rvmcode} & Matting with frame memory & ratio 0.8 or 0.4 & ONNX Runtime & 15.0 MB & GPL-3.0 \\
    7 & SINet~\cite{park2019sinet,sinetcode} & Segmentation & 224x224 & PyTorch & 0.44 MB & MIT \\
    8 & \utwonet{}, human~\cite{qin2020u2net,u2netcode} & Segmentation & 320x320 & PyTorch & 176.3 MB & Apache-2.0 \\
    9 & DeepLabV3 MobileNetV3~\cite{chen2017deeplabv3,howard2019mobilenetv3,torchvisiondeeplab} & General segmentation & s520 or s256 & PyTorch & 44.4 MB & BSD-3-Clause \\
    10 & BodyPix MobileNet 0.75~\cite{bodypix,tfbodypix} & Segmentation & half the frame & TensorFlow & 5.2 MB & Apache-2.0 \\
    11 & fast\_person\_segmentation~\cite{fastpersonseg,portraitseg} & Segmentation & 128x128 & OpenCV DNN & 14.7 MB & Unclear \\
    12 & ncnn ERDNet~\cite{ncnnportrait,fastportrait,ncnn} & Matting & 256x256 & ncnn & 3.3 MB & Unclear (model) \\
    13 & ERD seg-matting~\cite{fastportrait,mobilematting} & Matting & 256x256 & PyTorch & 3.4 MB & Unclear \\
    14 & SlimNet~\cite{portraitseg} & Segmentation & 512x512 & LiteRT & 1.5 MB & MIT \\
    15 & YOLO26n-seg, YOLO11n-seg, YOLOv8n-seg~\cite{ultralytics,yolo26} & Instance segmentation & l640 or l320 & Ultralytics & 6.7 MB & AGPL-3.0 \\
    \midrule
    \rowcolor{grp}\multicolumn{7}{@{}l}{\textbf{Complete pipelines} (our Python versions)} \\
    16 & Webinar HumanSeg~\cite{webinar} & Setup 2, temporal step, joint bilateral upsampling & 256x144 & ONNX Runtime & 3.8 MB & Model Apache-2.0, code unclear \\
    17 & Volcomix virtual-background~\cite{volcomix} & Meet~\cite{googlemeetblog} or ML Kit~\cite{mlkitselfie}, bilateral filter, light wrap & 160x96 to 256x256 & LiteRT & 0.25--0.4 MB & Apache-2.0, Meet models unclear \\
    18 & gregblur~\cite{gregblur} & MediaPipe selfie, bilateral filter, temporal blend & 256x256 & MediaPipe & 0.25 MB & Apache-2.0 \\
    19 & Linux-Fake-Background-Webcam~\cite{lfbw} & Setup 3, threshold, growth, box blur & 256x144 & MediaPipe & 0.25 MB & GPL-3.0 (project) \\
    20 & AI-Segmenter~\cite{aisegmenter} & RVM with YOLO11n boxes, or BiRefNet~\cite{zheng2024birefnet,birefnetcode} & ratio 0.25, l320, or 512x512 & ONNX Runtime, PyTorch & 21 MB or 445 MB & MIT app, uses GPL and AGPL \\
    \bottomrule
  \end{tabularx}
\end{table*}

A few details help to read the table. Setups 3 and 19 use the same model file, and setup 19 only adds a different clean up. Setups 12 and 13 are the same network. Setup 12 is a converted copy for the ncnn library, and in our test their masks differ by only 0.58 out of 255 on average. Setup 16 uses the same network as setup 2, converted to ONNX by the webinar project. For gregblur the default model is MediaPipe Selfie Multiclass, but we mainly report its faster selfie segmenter option.

We planned three more options but could not test them. The Deepixel Upperbody-Segmentation SDK~\cite{deepixel} is not publicly available, and its repository only has a README. Apple Vision person segmentation needs macOS, and NVIDIA Maxine needs an NVIDIA RTX GPU. Both were outside our CPU only setup on Windows.

\section{Experimental Setup}
\label{sec:setup}

\subsection{Hardware and software}
All runs used one laptop with an Intel Core i7-13620H CPU, 16 GB of RAM and Windows 11. The laptop also has an NVIDIA GeForce RTX 4050 Laptop GPU, but we did not use it. Every setup ran on the CPU only, with Python 3.10.0. The model files were downloaded from their official sources.

\subsection{Common program structure}
Every setup reads a frame, prepares it the way the model's authors describe, runs the model, applies the setup's own extra steps, and then produces the four outputs of \figurename~\ref{fig:example}. The plain models all use the same OpenCV blend to put the person on the new background. The pipelines use their own blending steps where the original project has them, for example light wrap in Volcomix and the halo free blur in gregblur.

\subsection{Inputs}
We used six inputs.
\begin{itemize}
  \item \emph{Test image.} A public domain NASA portrait of Kalpana Chawla in an orange suit, with a US flag behind her, at 960x1200. The flag is a hard case, because a busy background next to the person often leaks into the mask. The suit also has small patches and a badge.
  \item \emph{Hand photos.} Three free photos from Pexels of a person waving in a video call. The first (960x1440) is a close front camera view, where the face, shoulders and open hand fill the picture. In the second (960x1440) the open hand is small and far away. In the third (1200x800) the hand is raised beside the head in front of a busy room. All three test whether the fingers survive.
  \item \emph{Test clip.} 75 frames at 640x480, made by moving a crop window across the same portrait. It gives a repeatable video input with the same flag and suit.
  \item \emph{Webcam.} A Chicony USB2.0 camera at 640x480, with a person at a desk in an office. Most webcam runs used 150 frames.
\end{itemize}
All speed numbers come from the test clip and the webcam. The images in this paper come only from the four photos.

\subsection{Measurements}
We report two numbers. \emph{FPS} is the speed of the whole pipeline, from the moment a frame is read until all outputs are ready. \emph{Inference time} is the time of the model call in milliseconds. For most setups this is only the network. For YOLO it is the whole Ultralytics call. FPS is the number to use when comparing setups. In our experience 15~FPS already looks usable in a call and 30~FPS looks smooth. We call a setup \emph{real time} if the whole pipeline reaches at least 20~FPS on our laptop, which leaves some room for the camera, the encoder and the call software.

\subsection{Quality assessment}
We had no ground truth masks for our inputs, so we did not compute accuracy scores such as intersection over union. Instead we compared the masks visually. On the test image we looked at background leaks (the flag), holes in the body and the hair. On the hand photos we looked at the fingers. On the test clip and the webcam we also looked for flicker and trails. These are careful observations, but they are not a measured score. Section~\ref{sec:limits} discusses this.

\subsection{Run to run variation}
Laptop speed changes between runs because of turbo boost, heat and the way Windows moves work between fast and slow cores. In our runs the same setup could vary by up to two times. We plugged in the charger, closed heavy programs, ran each setup two or three times, and report the typical value or the range.

\section{Results}
\label{sec:results}

\subsection{Runtime performance}
\label{sec:speed}

\begin{figure}[t]
  \centering
  \begin{tikzpicture}
    \begin{axis}[
        xbar, bar width=4.4pt, bar shift=0pt,
        width=0.98\columnwidth, height=11.6cm,
        xmin=0, xmax=128,
        xlabel={Frames per second on the test clip},
        xlabel style={font=\footnotesize},
        ytick={1,...,29}, yticklabels={{BiRefNet (20)},{\utwonet{} (8)},{gregblur multiclass (18)},{DeepLabV3 520 (9)},{MODNet 512 (5)},{MP Multiclass (4)},{RVM ratio 0.8 (6)},{BodyPix (10)},{AI-Segmenter (20)},{DeepLabV3 256 (9)},{YOLO26 640 (15)},{YOLO11 640 (15)},{MODNet 256 (5)},{PP-HumanSeg v1 (1)},{YOLO26 320 (15)},{gregblur selfie (18)},{SINet (7)},{Volcomix Meet lite (17)},{PP-HumanSegV2 (2)},{RVM ratio 0.4 (6)},{Volcomix Meet full (17)},{Webinar HumanSeg (16)},{ERD PyTorch (13)},{SlimNet (14)},{fast\_person\_seg. (11)},{Volcomix ML Kit (17)},{ncnn ERDNet (12)},{LFBW recipe (19)},{MP Selfie (3)}},
        yticklabel style={font=\tiny},
        xticklabel style={font=\tiny},
        ymin=0.3, ymax=29.7,
        xmajorgrids, grid style={gray!25},
        axis x line*=bottom, axis y line*=left,
        nodes near coords, nodes near coords align={horizontal},
        every node near coord/.append style={font=\tiny, black!75},
        legend style={font=\tiny, at={(0.98,0.03)}, anchor=south east, draw=gray!40, fill=white},
        legend cell align=left,
      ]
      \addplot[fill=segc!75, draw=segc] coordinates {(1.9,2) (4.3,4) (7.2,6) (10.8,8) (13.7,10) (16.1,11) (18.4,12) (22.5,14) (22.9,15) (23.5,17) (26.5,19) (43.3,24) (43.7,25) (110,29)};
      \addplot[fill=matc!75, draw=matc] coordinates {(5.7,5) (10.0,7) (20.1,13) (27.3,20) (41.5,23) (47.1,27)};
      \addplot[fill=pipc!75, draw=pipc] coordinates {(0.1,1) (4.1,3) (13.5,9) (23.4,16) (24.6,18) (30.2,21) (33.9,22) (45.7,26) (65.8,28)};
      \legend{Segmentation model, Matting model, Complete pipeline}
      \draw[red!70!black, dashed, line width=0.7pt] (axis cs:20,0.3) -- (axis cs:20,29.7);
      \node[font=\tiny, red!70!black, anchor=south west, rotate=90] at (axis cs:21.5,0.6) {20 FPS};
    \end{axis}
  \end{tikzpicture}
  \caption{Whole pipeline speed on the test clip (middle of the measured range), coloured by the type of setup. The dashed line marks 20~FPS, our threshold for real time. Numbers in brackets are the setup numbers of Table~\ref{tab:setups}.}
  \label{fig:speedplot}
\end{figure}
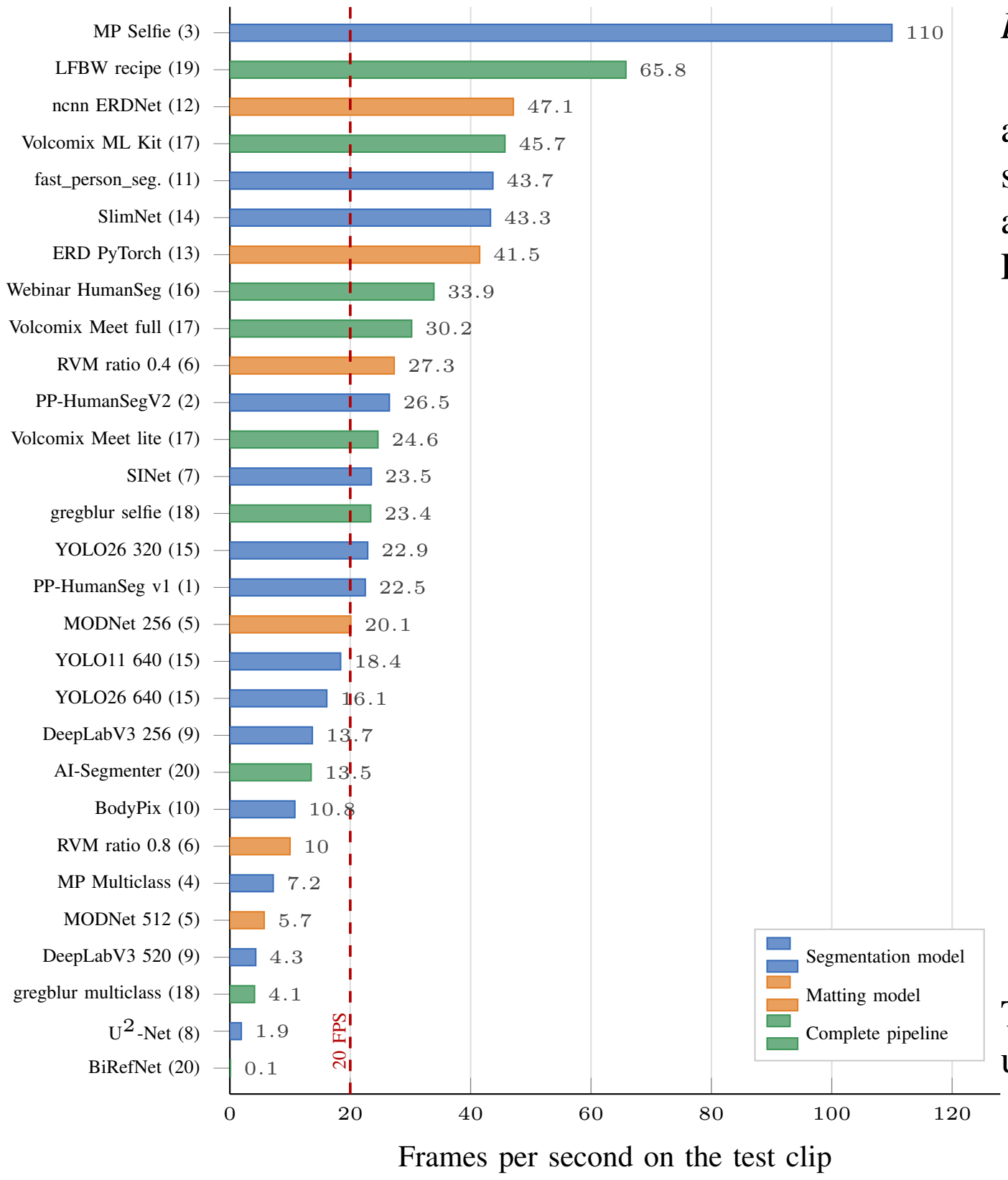

Table~\ref{tab:speed} lists the speed of every setup and \figurename~\ref{fig:speedplot} shows the same data as a chart.

\begin{table*}[!t]
  \centering
  \caption{Speed on an Intel Core i7-13620H laptop CPU (no GPU) at 640x480. FPS is for the whole pipeline. Inference is the model call in milliseconds, given for video and webcam where both were measured. A range covers several runs and a dash means not measured. Setups are sorted by video FPS.}
  \label{tab:speed}
  \footnotesize
  \renewcommand{\arraystretch}{1.12}
  \begin{tabular}{@{}r L{6.4cm} C{2.3cm} C{2.3cm} C{2.8cm}@{}}
    \toprule
    \textbf{\#} & \textbf{Setup and setting} & \textbf{FPS (video)} & \textbf{FPS (webcam)} & \textbf{Inference (ms)} \\
    \midrule
    \rowcolor{grp}\multicolumn{5}{@{}l}{\textbf{40 FPS and more}} \\
    3 & MediaPipe Selfie, landscape & 110 & 41--63 & 3.6 / 6--11 \\
    19 & Linux-Fake-Background-Webcam & 63.2--68.4 & 53.3--53.4 & 5.8--6.2 / 7.4--8.1 \\
    12 & ncnn ERDNet, 4 threads & 47.1 & 38.1--42.9 & 15.2 / 17.1--18.9 \\
    17 & Volcomix with ML Kit selfie & 44.6--46.7 & 41.8 & 4.8--4.9 \\
    11 & fast\_person\_segmentation & 43.7 & 21.7--23.4 & 14.2 / 26.6--27.9 \\
    14 & SlimNet, 2 threads & 41.0--45.5 & 15.2--16.6 & 12.7--13.3 / 39--44 \\
    13 & ERD seg-matting, PyTorch & 41.5 & 36.5 & 17.1 / 19.5 \\
    \midrule
    \rowcolor{grp}\multicolumn{5}{@{}l}{\textbf{20 to 40 FPS}} \\
    16 & Webinar HumanSeg (network only in inference) & 33.0--34.8 & 34.2--34.6 & 4.9--5.3 \\
    17 & Volcomix with Google Meet full & 29.9--30.4 & 44.4 & 2.8--3.0 \\
    6 & RVM, ratio 0.4 & \multicolumn{2}{c}{25.3--29.3 (both)} & 25--29 \\
    2 & PP-HumanSegV2 Lite & 26--27 & -- & 32--34 \\
    17 & Volcomix with Google Meet lite & 24.3--24.8 & 21.0--25.6 & 1.3 / 2.0--2.6 \\
    7 & SINet & \multicolumn{2}{c}{22.7--24.3 (both)} & 35--37 \\
    18 & gregblur with selfie segmenter & 23.3--23.5 & 17.7--19.5 & 6.1--6.2 \\
    15 & YOLO26n-seg, input 320 & 22.9 & 32.1 & 35 / 25 \\
    1 & PP-HumanSeg v1 & 22--23 & 20.3 & 37--39 / 41.5 \\
    5 & MODNet, input 256 & \multicolumn{2}{c}{19.6--20.5 (both)} & 40--43 \\
    \midrule
    \rowcolor{grp}\multicolumn{5}{@{}l}{\textbf{10 to 20 FPS}} \\
    15 & YOLO11n-seg / YOLOv8n-seg, input 640 & 18.4 / 17.9 & -- & 47.7 / 49.6 \\
    15 & YOLO26n-seg, input 640 & 13.7--18.5 & 16.1--16.8 & 47--66 / 53--55 \\
    20 & AI-Segmenter, RVM and YOLO11n & 12.3--14.6 & -- & 60--66 \\
    9 & DeepLabV3, input 256 & \multicolumn{2}{c}{13.4--13.9 (both)} & 54--55 \\
    10 & BodyPix & \multicolumn{2}{c}{10.5--11.1 (both)} & 63--66 \\
    6 & RVM, default ratio 0.8 & \multicolumn{2}{c}{9.9--10.1 (both)} & 88 \\
    \midrule
    \rowcolor{grp}\multicolumn{5}{@{}l}{\textbf{Below 10 FPS}} \\
    4 & MediaPipe Selfie Multiclass & 7.2 & 8.3 & 133 / 116 \\
    5 & MODNet, input 512 & \multicolumn{2}{c}{5.6--5.8 (both)} & 161--166 \\
    9 & DeepLabV3, input 520 & \multicolumn{2}{c}{4.3 (both)} & 162--176 \\
    18 & gregblur with Selfie Multiclass (its default) & 3.8--4.3 & 4.9 & 114--117 \\
    8 & \utwonet{} & \multicolumn{2}{c}{1.9 (both)} & 518--523 \\
    20 & AI-Segmenter with BiRefNet & 0.1 & -- & 9{,}000--23{,}000 \\
    \bottomrule
  \end{tabular}
\end{table*}

Seven setups reach 40~FPS or more on the test clip. All of them use a small input, between 128x128 and 512x512, and a small network. The webcam can be much slower than the video file, probably because reading the camera also uses CPU time. SlimNet dropped from about 43~FPS to about 16~FPS and fast\_person\_segmentation from about 44 to about 23. For a few setups the webcam number was higher than the video number, which shows how much the laptop speed moves between runs.

Matting models are slow at their default input size but become real time, or close to it, at a smaller one. MODNet goes from about 6~FPS at 512 to about 20~FPS at 256, and RVM goes from about 10~FPS at ratio 0.8 to 25--29~FPS at ratio 0.4. \utwonet{}, BiRefNet, DeepLabV3 at full size and MediaPipe Multiclass are not real time on this CPU. The AI-Segmenter numbers are less reliable than the others, because other programs used about half of the CPU during that test.

\subsection{Mask quality on the test image}
\label{sec:quality}

\begin{figure*}[!t]
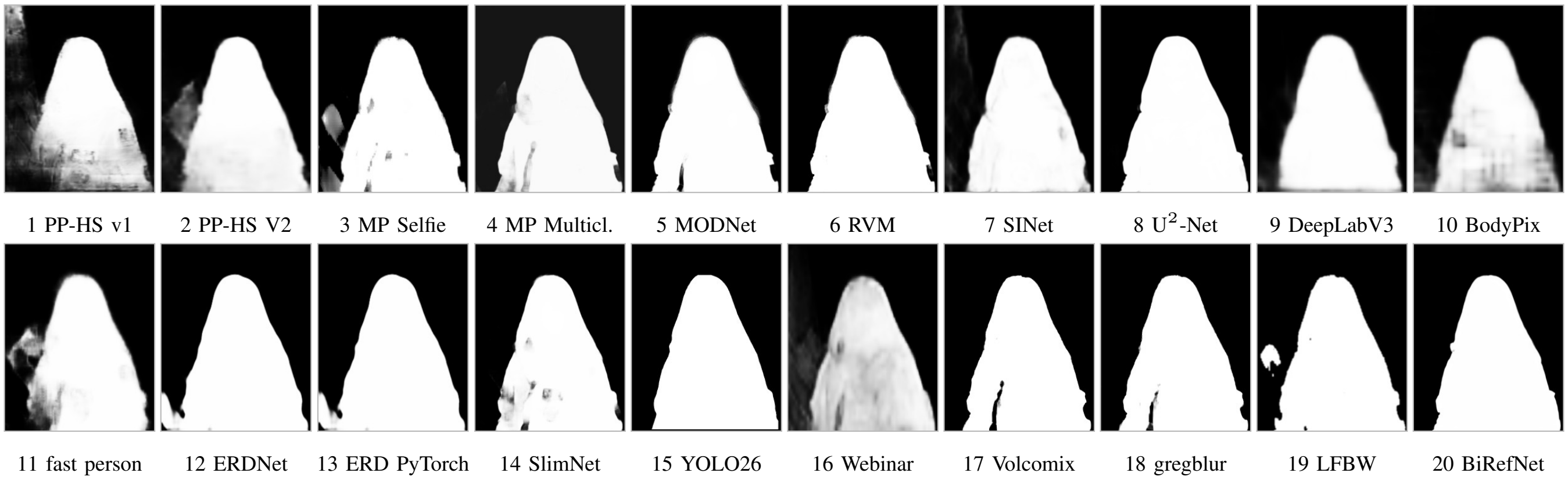

  \centering
  \setlength{\tabcolsep}{0.8pt}
  \begin{tabular}{cccccccccc}
    \tile{0.0935\textwidth}{masks/pphsv1.jpg}{1 PP-HS v1} &
    \tile{0.0935\textwidth}{masks/pphsv2.jpg}{2 PP-HS V2} &
    \tile{0.0935\textwidth}{masks/mpselfie.jpg}{3 MP Selfie} &
    \tile{0.0935\textwidth}{masks/mpmulti.jpg}{4 MP Multicl.} &
    \tile{0.0935\textwidth}{masks/modnet512.jpg}{5 MODNet} &
    \tile{0.0935\textwidth}{masks/rvm.jpg}{6 RVM} &
    \tile{0.0935\textwidth}{masks/sinet.jpg}{7 SINet} &
    \tile{0.0935\textwidth}{masks/u2net.jpg}{8 \utwonet{}} &
    \tile{0.0935\textwidth}{masks/deeplab.jpg}{9 DeepLabV3} &
    \tile{0.0935\textwidth}{masks/bodypix.jpg}{10 BodyPix} \\[3pt]
    \tile{0.0935\textwidth}{masks/fastperson.jpg}{11 fast person} &
    \tile{0.0935\textwidth}{masks/erdnet.jpg}{12 ERDNet} &
    \tile{0.0935\textwidth}{masks/erdtorch.jpg}{13 ERD PyTorch} &
    \tile{0.0935\textwidth}{masks/slimnet.jpg}{14 SlimNet} &
    \tile{0.0935\textwidth}{masks/yolo26.jpg}{15 YOLO26} &
    \tile{0.0935\textwidth}{masks/webinar.jpg}{16 Webinar} &
    \tile{0.0935\textwidth}{masks/volc_mlkit.jpg}{17 Volcomix} &
    \tile{0.0935\textwidth}{masks/gregblur.jpg}{18 gregblur} &
    \tile{0.0935\textwidth}{masks/lfbw.jpg}{19 LFBW} &
    \tile{0.0935\textwidth}{masks/aiseg_biref.jpg}{20 BiRefNet}
  \end{tabular}
  \caption{Masks of all 20 setups on the test image (white is person). Setup 20 is shown with BiRefNet. MODNet is shown at input 512, YOLO26 at input 640, Volcomix with the ML Kit model and gregblur with the selfie segmenter.}
  \label{fig:masks}
\end{figure*}

\begin{figure*}[!t]
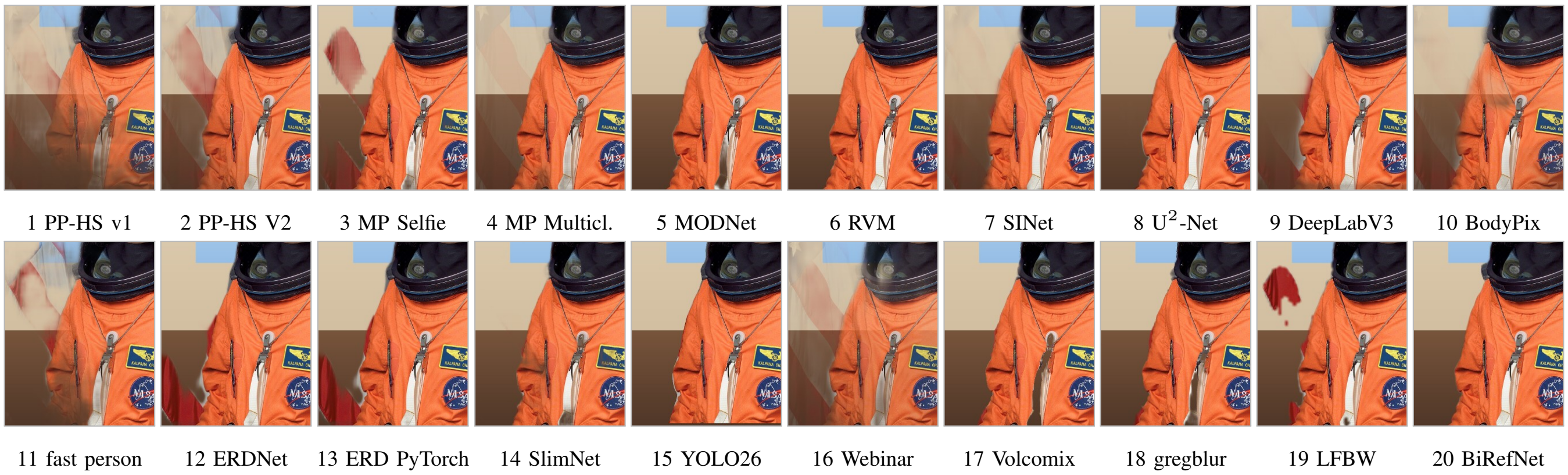

  \centering
  \setlength{\tabcolsep}{0.8pt}
  \begin{tabular}{cccccccccc}
    \tile{0.0935\textwidth}{flag/pphsv1.jpg}{1 PP-HS v1} &
    \tile{0.0935\textwidth}{flag/pphsv2.jpg}{2 PP-HS V2} &
    \tile{0.0935\textwidth}{flag/mpselfie.jpg}{3 MP Selfie} &
    \tile{0.0935\textwidth}{flag/mpmulti.jpg}{4 MP Multicl.} &
    \tile{0.0935\textwidth}{flag/modnet512.jpg}{5 MODNet} &
    \tile{0.0935\textwidth}{flag/rvm.jpg}{6 RVM} &
    \tile{0.0935\textwidth}{flag/sinet.jpg}{7 SINet} &
    \tile{0.0935\textwidth}{flag/u2net.jpg}{8 \utwonet{}} &
    \tile{0.0935\textwidth}{flag/deeplab.jpg}{9 DeepLabV3} &
    \tile{0.0935\textwidth}{flag/bodypix.jpg}{10 BodyPix} \\[3pt]
    \tile{0.0935\textwidth}{flag/fastperson.jpg}{11 fast person} &
    \tile{0.0935\textwidth}{flag/erdnet.jpg}{12 ERDNet} &
    \tile{0.0935\textwidth}{flag/erdtorch.jpg}{13 ERD PyTorch} &
    \tile{0.0935\textwidth}{flag/slimnet.jpg}{14 SlimNet} &
    \tile{0.0935\textwidth}{flag/yolo26.jpg}{15 YOLO26} &
    \tile{0.0935\textwidth}{flag/webinar.jpg}{16 Webinar} &
    \tile{0.0935\textwidth}{flag/volc_mlkit.jpg}{17 Volcomix} &
    \tile{0.0935\textwidth}{flag/gregblur.jpg}{18 gregblur} &
    \tile{0.0935\textwidth}{flag/lfbw.jpg}{19 LFBW} &
    \tile{0.0935\textwidth}{flag/aiseg_biref.jpg}{20 BiRefNet}
  \end{tabular}
  \caption{Close view of the lower left of the replaced background, next to the flag, for all 20 setups. A solid patch of the flag stays with MediaPipe Selfie (3) and LFBW (19). See through stripes appear with PP-HumanSeg v1 (1), BodyPix (10), fast\_person\_segmentation (11) and the Webinar pipeline (16), and fainter ones with 2, 4 and 7. ERDNet (12, 13) keeps one stripe next to the arm, and Volcomix (17) and gregblur (18) show a little of the flag in a gap near the arm. MODNet, RVM, \utwonet{}, SlimNet, YOLO26 and BiRefNet show no visible leak. Setup 20 is shown with BiRefNet. MODNet is shown at input 512, YOLO26 at input 640, Volcomix with the ML Kit model and gregblur with the selfie segmenter.}
  \label{fig:flag}
\end{figure*}

\begin{figure*}[!t]
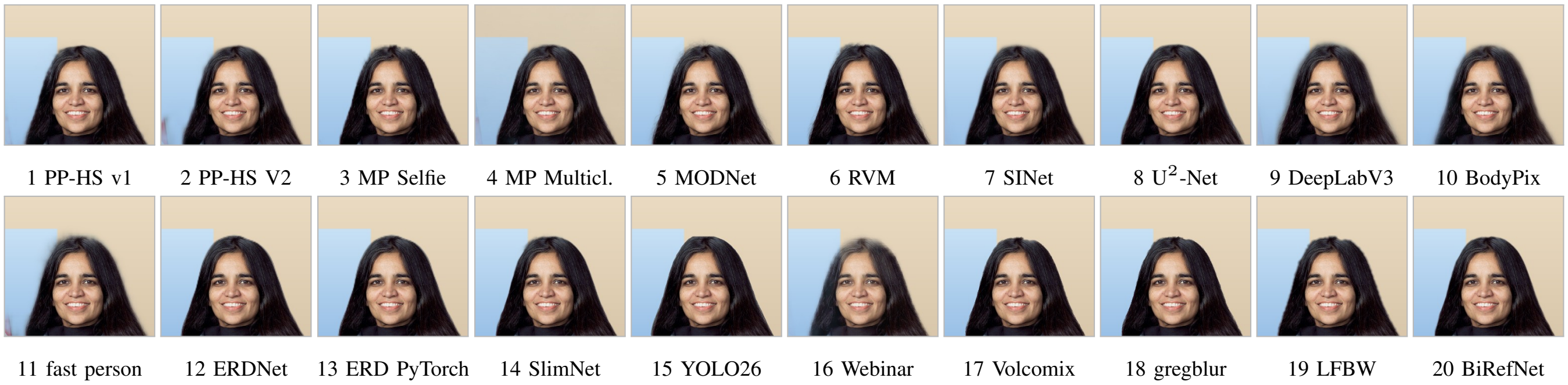

  \centering
  \setlength{\tabcolsep}{0.8pt}
  \begin{tabular}{cccccccccc}
    \tile{0.0935\textwidth}{hair/pphsv1.jpg}{1 PP-HS v1} &
    \tile{0.0935\textwidth}{hair/pphsv2.jpg}{2 PP-HS V2} &
    \tile{0.0935\textwidth}{hair/mpselfie.jpg}{3 MP Selfie} &
    \tile{0.0935\textwidth}{hair/mpmulti.jpg}{4 MP Multicl.} &
    \tile{0.0935\textwidth}{hair/modnet512.jpg}{5 MODNet} &
    \tile{0.0935\textwidth}{hair/rvm.jpg}{6 RVM} &
    \tile{0.0935\textwidth}{hair/sinet.jpg}{7 SINet} &
    \tile{0.0935\textwidth}{hair/u2net.jpg}{8 \utwonet{}} &
    \tile{0.0935\textwidth}{hair/deeplab.jpg}{9 DeepLabV3} &
    \tile{0.0935\textwidth}{hair/bodypix.jpg}{10 BodyPix} \\[3pt]
    \tile{0.0935\textwidth}{hair/fastperson.jpg}{11 fast person} &
    \tile{0.0935\textwidth}{hair/erdnet.jpg}{12 ERDNet} &
    \tile{0.0935\textwidth}{hair/erdtorch.jpg}{13 ERD PyTorch} &
    \tile{0.0935\textwidth}{hair/slimnet.jpg}{14 SlimNet} &
    \tile{0.0935\textwidth}{hair/yolo26.jpg}{15 YOLO26} &
    \tile{0.0935\textwidth}{hair/webinar.jpg}{16 Webinar} &
    \tile{0.0935\textwidth}{hair/volc_mlkit.jpg}{17 Volcomix} &
    \tile{0.0935\textwidth}{hair/gregblur.jpg}{18 gregblur} &
    \tile{0.0935\textwidth}{hair/lfbw.jpg}{19 LFBW} &
    \tile{0.0935\textwidth}{hair/aiseg_biref.jpg}{20 BiRefNet}
  \end{tabular}
  \caption{Close view of the hair on the replaced background for all 20 setups. The matting models MODNet (5) and RVM (6) keep loose strands, and MODNet also carries a thin dark fringe of the old background. YOLO26 (15) cuts the hair with a hard edge. DeepLabV3 (9), fast\_person\_segmentation (11) and the Webinar pipeline (16) leave a dark haze around the hair, and MediaPipe Multiclass (4) tints the whole background slightly. Setup 20 is shown with BiRefNet. MODNet is shown at input 512, YOLO26 at input 640, Volcomix with the ML Kit model and gregblur with the selfie segmenter.}
  \label{fig:hair}
\end{figure*}

\figurename~\ref{fig:masks} shows the mask of every setup on the test image. \figurename~\ref{fig:flag} zooms into the flag side of the replaced background, where most leaks appear, and \figurename~\ref{fig:hair} zooms into the hair. Table~\ref{tab:quality} sums up what we saw. It also lists the notes we made on the test clip and the webcam, where temporal behaviour can be seen. The frame memory of RVM (6) and the temporal steps of setups 16 and 18 cannot help on a single image, so the photos show them at their weakest.

\begin{table*}[!t]
  \centering
  \caption{Observations on mask quality. The first column refers to the test image (Figs.~\ref{fig:masks}, \ref{fig:flag} and \ref{fig:hair}). The second column holds notes from the test clip and the webcam.}
  \label{tab:quality}
  \footnotesize
  \renewcommand{\arraystretch}{1.15}
  \setlength{\tabcolsep}{4pt}
  \begin{tabularx}{\textwidth}{@{}r p{3.6cm} Y Y@{}}
    \toprule
    \textbf{\#} & \textbf{Setup} & \textbf{Test image} & \textbf{Test clip and webcam} \\
    \midrule
    1 & PP-HumanSeg v1 & See through stripes of the flag leak in over a large area. & Soft edges. \\
    2 & PP-HumanSegV2 Lite & Faint stripes of the flag. The body is more solid than in v1. & Cleaner than v1. \\
    3 & MediaPipe Selfie & A solid red patch of the flag is kept. Sharp outline. & Sharp outline, small leaks of busy backgrounds. \\
    4 & MediaPipe Multiclass & The flag is mostly gone, with faint stripes left, and the whole background has a grey value of about 8\%. & One of the cleanest masks, but slow. \\
    5 & MODNet & No flag leak. Loose hair strands are kept, with a thin dark fringe. & One of the best mattes. \\
    6 & RVM & No flag leak. Loose hair strands are kept. & As clean as MODNet and steady from frame to frame. \\
    7 & SINet & One faint streak of the flag. & Edges a little soft. \\
    8 & \utwonet{} & Clean and smooth outline. & Little hair detail. \\
    9 & DeepLabV3 & No clear flag leak, but a soft and wide edge with a dark haze around the hair. & Blobby edges with a wide halo. \\
    10 & BodyPix & See through leak and grey patches inside the body. & The weakest mask, with blocky edges. \\
    11 & fast\_person\_segmentation & Clear stripes of the flag leak in, and a dark haze around the hair. & Solid body, soft edges. \\
    12 & ncnn ERDNet & Clean body and smooth outline, but one red stripe of the flag next to the arm is kept. & Very clean and smooth, close to MODNet. \\
    13 & ERD seg-matting, PyTorch & Same as setup 12, including the stripe next to the arm. & Same as setup 12. \\
    14 & SlimNet & No flag leak, but several small holes in the suit. & Small holes can appear, some flicker. \\
    15 & YOLO26n-seg & No flag leak and no holes. Hard outline that cuts the hair. & Same at input 320. YOLO11n had holes at the shoulder patch and the badge. Separates people. \\
    16 & Webinar HumanSeg & Soft mask with faint flag stripes and a dark haze around the hair. & Edge follows hair and shoulders more closely than setup 2, less flicker. \\
    17 & Volcomix, ML Kit & Clean, with a small gap near the left arm where the flag shows. & Cleanest of its three models. Both Meet models left a small hole at the shoulder patch. \\
    18 & gregblur & Clean, with a small gap near the left arm where the flag shows. & Removes small specks, blur without halo. \\
    19 & \mbox{Linux-Fake-Background-}\allowbreak Webcam & A solid patch of the flag is kept. & The hard threshold keeps the flag patch solid, a few small holes. \\
    20 & AI-Segmenter & RVM version clean. BiRefNet version gives the cleanest edge of all. & The YOLO boxes did not remove leaks inside the person's box. \\
    \bottomrule
  \end{tabularx}
\end{table*}

Two more observations came from live webcam use. First, in live use of the Webinar HumanSeg pipeline, a hand held up close to the camera was usually kept, but a hand further away became small in the model input. With a 256x144 input a finger is only about two model pixels wide, and the fingers could disappear. Section~\ref{sec:hands} looks at hands in more detail. Second, the plain segmentation and matting models see everyone in the frame as one person, so a colleague sitting behind the speaker appears in the mask too. YOLO is the only model in our test that gives a separate mask for each person, so it can keep only the biggest one. The AI-Segmenter pipeline also uses YOLO, but only to find boxes around all people, so it keeps everyone.

\subsection{Hands and fingers in video call photos}
\label{sec:hands}

\begin{figure}[t]
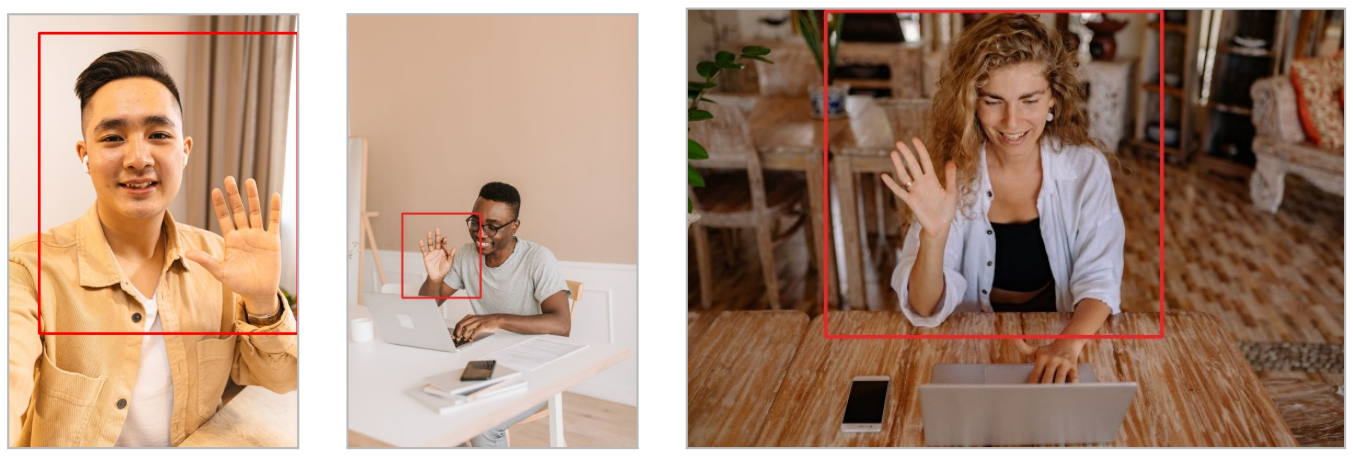

  \centering
  \tile{0.215\columnwidth}{hand0/input.jpg}{(a) Front camera}\hfill
  \tile{0.215\columnwidth}{hand/input.jpg}{(b) Far hand}\hfill
  \tile{0.49\columnwidth}{hand2/input.jpg}{(c) Close hand}
  \caption{The three video call photos. The red boxes mark the regions shown in Figs.~\ref{fig:handmasks0}, \ref{fig:handmasks} and~\ref{fig:handmasks2}. Photos by ANTONI SHKRABA production (a), Vlada Karpovich (b) and Yan Krukau (c) on Pexels.}
  \label{fig:handinput}
\end{figure}

\begin{figure*}[!t]
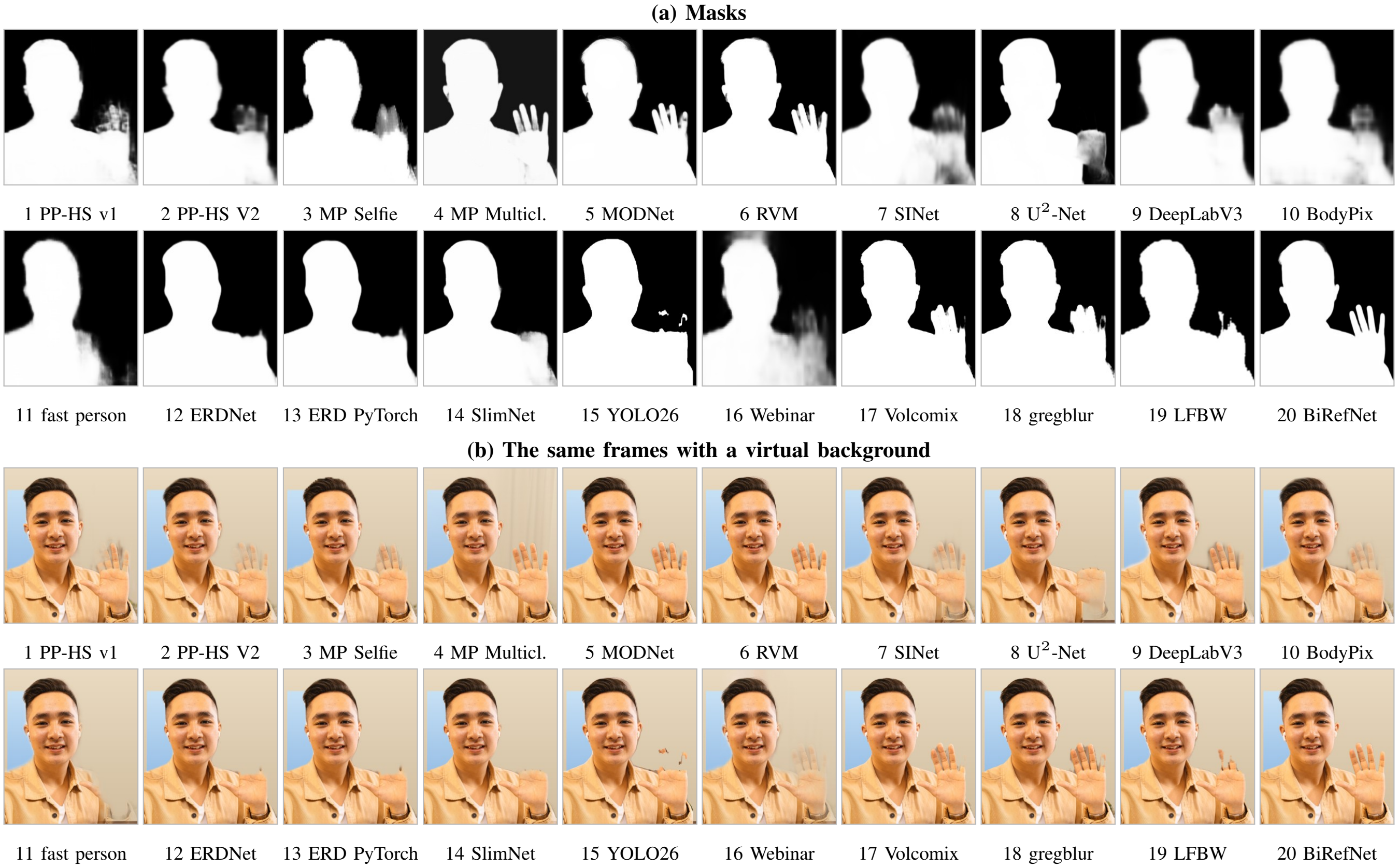

  \centering
  {\footnotesize\textbf{(a) Masks}}\\[2pt]
  \setlength{\tabcolsep}{0.8pt}
  \begin{tabular}{cccccccccc}
    \tile{0.0935\textwidth}{hand0/masks/pphsv1.jpg}{1 PP-HS v1} &
    \tile{0.0935\textwidth}{hand0/masks/pphsv2.jpg}{2 PP-HS V2} &
    \tile{0.0935\textwidth}{hand0/masks/mpselfie.jpg}{3 MP Selfie} &
    \tile{0.0935\textwidth}{hand0/masks/mpmulti.jpg}{4 MP Multicl.} &
    \tile{0.0935\textwidth}{hand0/masks/modnet512.jpg}{5 MODNet} &
    \tile{0.0935\textwidth}{hand0/masks/rvm.jpg}{6 RVM} &
    \tile{0.0935\textwidth}{hand0/masks/sinet.jpg}{7 SINet} &
    \tile{0.0935\textwidth}{hand0/masks/u2net.jpg}{8 \utwonet{}} &
    \tile{0.0935\textwidth}{hand0/masks/deeplab.jpg}{9 DeepLabV3} &
    \tile{0.0935\textwidth}{hand0/masks/bodypix.jpg}{10 BodyPix} \\[3pt]
    \tile{0.0935\textwidth}{hand0/masks/fastperson.jpg}{11 fast person} &
    \tile{0.0935\textwidth}{hand0/masks/erdnet.jpg}{12 ERDNet} &
    \tile{0.0935\textwidth}{hand0/masks/erdtorch.jpg}{13 ERD PyTorch} &
    \tile{0.0935\textwidth}{hand0/masks/slimnet.jpg}{14 SlimNet} &
    \tile{0.0935\textwidth}{hand0/masks/yolo26.jpg}{15 YOLO26} &
    \tile{0.0935\textwidth}{hand0/masks/webinar.jpg}{16 Webinar} &
    \tile{0.0935\textwidth}{hand0/masks/volc_mlkit.jpg}{17 Volcomix} &
    \tile{0.0935\textwidth}{hand0/masks/gregblur.jpg}{18 gregblur} &
    \tile{0.0935\textwidth}{hand0/masks/lfbw.jpg}{19 LFBW} &
    \tile{0.0935\textwidth}{hand0/masks/aiseg_biref.jpg}{20 BiRefNet}
  \end{tabular}
  \\[4pt]
  {\footnotesize\textbf{(b) The same frames with a virtual background}}\\[2pt]
  \setlength{\tabcolsep}{0.8pt}
  \begin{tabular}{cccccccccc}
    \tile{0.0935\textwidth}{hand0/replace/pphsv1.jpg}{1 PP-HS v1} &
    \tile{0.0935\textwidth}{hand0/replace/pphsv2.jpg}{2 PP-HS V2} &
    \tile{0.0935\textwidth}{hand0/replace/mpselfie.jpg}{3 MP Selfie} &
    \tile{0.0935\textwidth}{hand0/replace/mpmulti.jpg}{4 MP Multicl.} &
    \tile{0.0935\textwidth}{hand0/replace/modnet512.jpg}{5 MODNet} &
    \tile{0.0935\textwidth}{hand0/replace/rvm.jpg}{6 RVM} &
    \tile{0.0935\textwidth}{hand0/replace/sinet.jpg}{7 SINet} &
    \tile{0.0935\textwidth}{hand0/replace/u2net.jpg}{8 \utwonet{}} &
    \tile{0.0935\textwidth}{hand0/replace/deeplab.jpg}{9 DeepLabV3} &
    \tile{0.0935\textwidth}{hand0/replace/bodypix.jpg}{10 BodyPix} \\[3pt]
    \tile{0.0935\textwidth}{hand0/replace/fastperson.jpg}{11 fast person} &
    \tile{0.0935\textwidth}{hand0/replace/erdnet.jpg}{12 ERDNet} &
    \tile{0.0935\textwidth}{hand0/replace/erdtorch.jpg}{13 ERD PyTorch} &
    \tile{0.0935\textwidth}{hand0/replace/slimnet.jpg}{14 SlimNet} &
    \tile{0.0935\textwidth}{hand0/replace/yolo26.jpg}{15 YOLO26} &
    \tile{0.0935\textwidth}{hand0/replace/webinar.jpg}{16 Webinar} &
    \tile{0.0935\textwidth}{hand0/replace/volc_mlkit.jpg}{17 Volcomix} &
    \tile{0.0935\textwidth}{hand0/replace/gregblur.jpg}{18 gregblur} &
    \tile{0.0935\textwidth}{hand0/replace/lfbw.jpg}{19 LFBW} &
    \tile{0.0935\textwidth}{hand0/replace/aiseg_biref.jpg}{20 BiRefNet}
  \end{tabular}
  \caption{The front camera photo (\figurename~\ref{fig:handinput}a), showing the face, shoulders and waving hand for all 20 setups, as a mask (a) and on a virtual background (b). MODNet (5), RVM (6) and BiRefNet (20) keep separate fingers, and MediaPipe Multiclass (4) shows them on a grey background. Volcomix (17) and gregblur (18) keep the fingers, but some gaps between them are filled. Seven setups give a blurry, blocky or noisy hand that is partly see through, \utwonet{} (8) turns the hand into a grey patch, and six setups lose the hand. MODNet is shown at input 512, YOLO26 at input 640, Volcomix with the ML Kit model, gregblur with the selfie segmenter and setup 20 with BiRefNet.}
  \label{fig:handmasks0}
\end{figure*}

\begin{figure*}[!t]
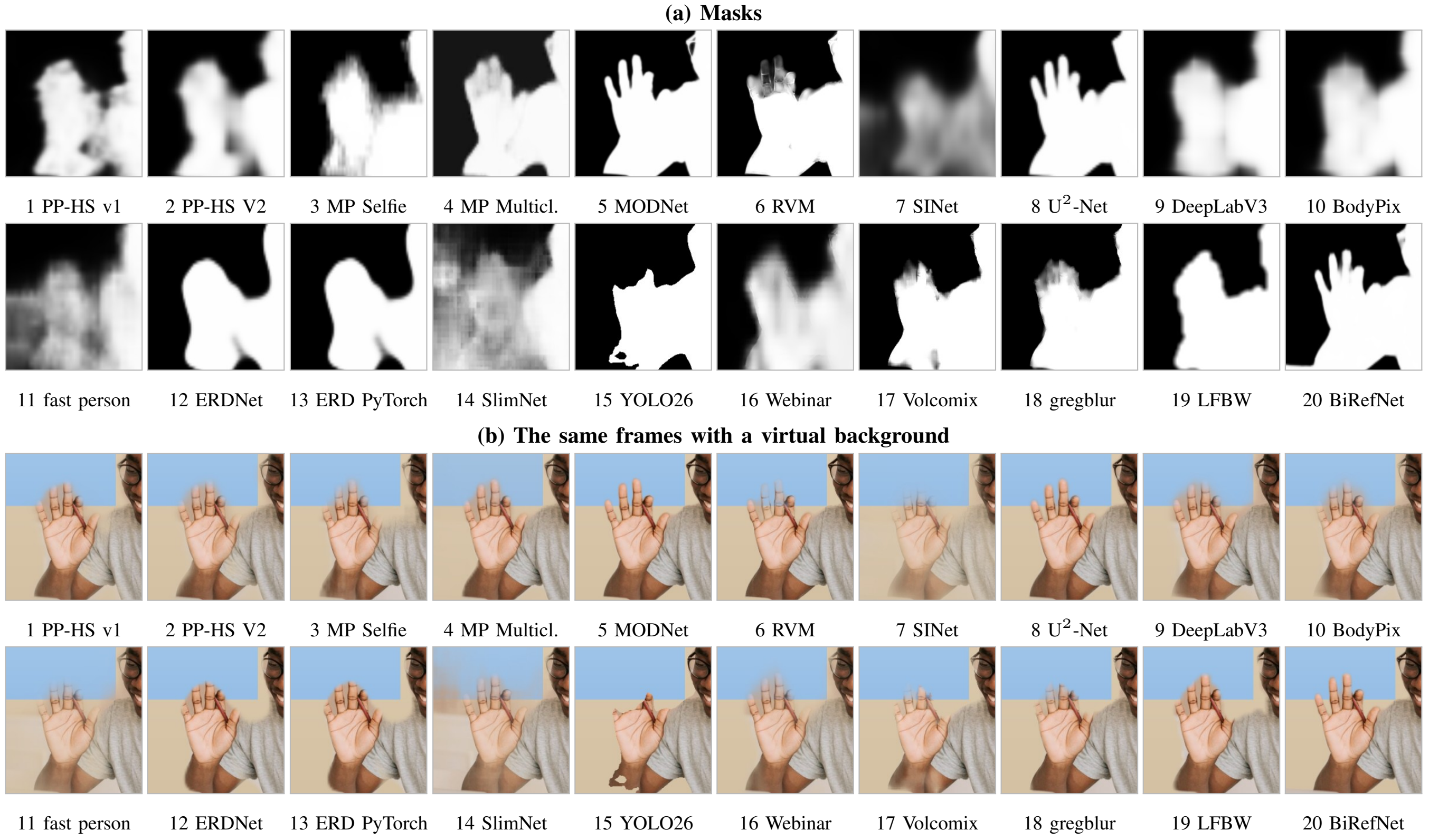

  \centering
  {\footnotesize\textbf{(a) Masks}}\\[2pt]
  \setlength{\tabcolsep}{0.8pt}
  \begin{tabular}{cccccccccc}
    \tile{0.0935\textwidth}{hand/masks/pphsv1.jpg}{1 PP-HS v1} &
    \tile{0.0935\textwidth}{hand/masks/pphsv2.jpg}{2 PP-HS V2} &
    \tile{0.0935\textwidth}{hand/masks/mpselfie.jpg}{3 MP Selfie} &
    \tile{0.0935\textwidth}{hand/masks/mpmulti.jpg}{4 MP Multicl.} &
    \tile{0.0935\textwidth}{hand/masks/modnet512.jpg}{5 MODNet} &
    \tile{0.0935\textwidth}{hand/masks/rvm.jpg}{6 RVM} &
    \tile{0.0935\textwidth}{hand/masks/sinet.jpg}{7 SINet} &
    \tile{0.0935\textwidth}{hand/masks/u2net.jpg}{8 \utwonet{}} &
    \tile{0.0935\textwidth}{hand/masks/deeplab.jpg}{9 DeepLabV3} &
    \tile{0.0935\textwidth}{hand/masks/bodypix.jpg}{10 BodyPix} \\[3pt]
    \tile{0.0935\textwidth}{hand/masks/fastperson.jpg}{11 fast person} &
    \tile{0.0935\textwidth}{hand/masks/erdnet.jpg}{12 ERDNet} &
    \tile{0.0935\textwidth}{hand/masks/erdtorch.jpg}{13 ERD PyTorch} &
    \tile{0.0935\textwidth}{hand/masks/slimnet.jpg}{14 SlimNet} &
    \tile{0.0935\textwidth}{hand/masks/yolo26.jpg}{15 YOLO26} &
    \tile{0.0935\textwidth}{hand/masks/webinar.jpg}{16 Webinar} &
    \tile{0.0935\textwidth}{hand/masks/volc_mlkit.jpg}{17 Volcomix} &
    \tile{0.0935\textwidth}{hand/masks/gregblur.jpg}{18 gregblur} &
    \tile{0.0935\textwidth}{hand/masks/lfbw.jpg}{19 LFBW} &
    \tile{0.0935\textwidth}{hand/masks/aiseg_biref.jpg}{20 BiRefNet}
  \end{tabular}
  \\[4pt]
  {\footnotesize\textbf{(b) The same frames with a virtual background}}\\[2pt]
  \setlength{\tabcolsep}{0.8pt}
  \begin{tabular}{cccccccccc}
    \tile{0.0935\textwidth}{hand/replace/pphsv1.jpg}{1 PP-HS v1} &
    \tile{0.0935\textwidth}{hand/replace/pphsv2.jpg}{2 PP-HS V2} &
    \tile{0.0935\textwidth}{hand/replace/mpselfie.jpg}{3 MP Selfie} &
    \tile{0.0935\textwidth}{hand/replace/mpmulti.jpg}{4 MP Multicl.} &
    \tile{0.0935\textwidth}{hand/replace/modnet512.jpg}{5 MODNet} &
    \tile{0.0935\textwidth}{hand/replace/rvm.jpg}{6 RVM} &
    \tile{0.0935\textwidth}{hand/replace/sinet.jpg}{7 SINet} &
    \tile{0.0935\textwidth}{hand/replace/u2net.jpg}{8 \utwonet{}} &
    \tile{0.0935\textwidth}{hand/replace/deeplab.jpg}{9 DeepLabV3} &
    \tile{0.0935\textwidth}{hand/replace/bodypix.jpg}{10 BodyPix} \\[3pt]
    \tile{0.0935\textwidth}{hand/replace/fastperson.jpg}{11 fast person} &
    \tile{0.0935\textwidth}{hand/replace/erdnet.jpg}{12 ERDNet} &
    \tile{0.0935\textwidth}{hand/replace/erdtorch.jpg}{13 ERD PyTorch} &
    \tile{0.0935\textwidth}{hand/replace/slimnet.jpg}{14 SlimNet} &
    \tile{0.0935\textwidth}{hand/replace/yolo26.jpg}{15 YOLO26} &
    \tile{0.0935\textwidth}{hand/replace/webinar.jpg}{16 Webinar} &
    \tile{0.0935\textwidth}{hand/replace/volc_mlkit.jpg}{17 Volcomix} &
    \tile{0.0935\textwidth}{hand/replace/gregblur.jpg}{18 gregblur} &
    \tile{0.0935\textwidth}{hand/replace/lfbw.jpg}{19 LFBW} &
    \tile{0.0935\textwidth}{hand/replace/aiseg_biref.jpg}{20 BiRefNet}
  \end{tabular}
  \caption{The far hand (\figurename~\ref{fig:handinput}b) for all 20 setups, as a mask (a) and on a virtual background (b). MODNet (5), \utwonet{} (8) and BiRefNet (20) keep separate fingers. MediaPipe Multiclass (4), RVM (6), Volcomix (17) and gregblur (18) show the fingers, but soft or partly see through. Most other setups turn the hand into one blob, three fade it into grey, and YOLO26 (15) cuts the fingers off. MODNet is shown at input 512, YOLO26 at input 640, Volcomix with the ML Kit model, gregblur with the selfie segmenter and setup 20 with BiRefNet.}
  \label{fig:handmasks}
\end{figure*}

\begin{figure*}[!t]
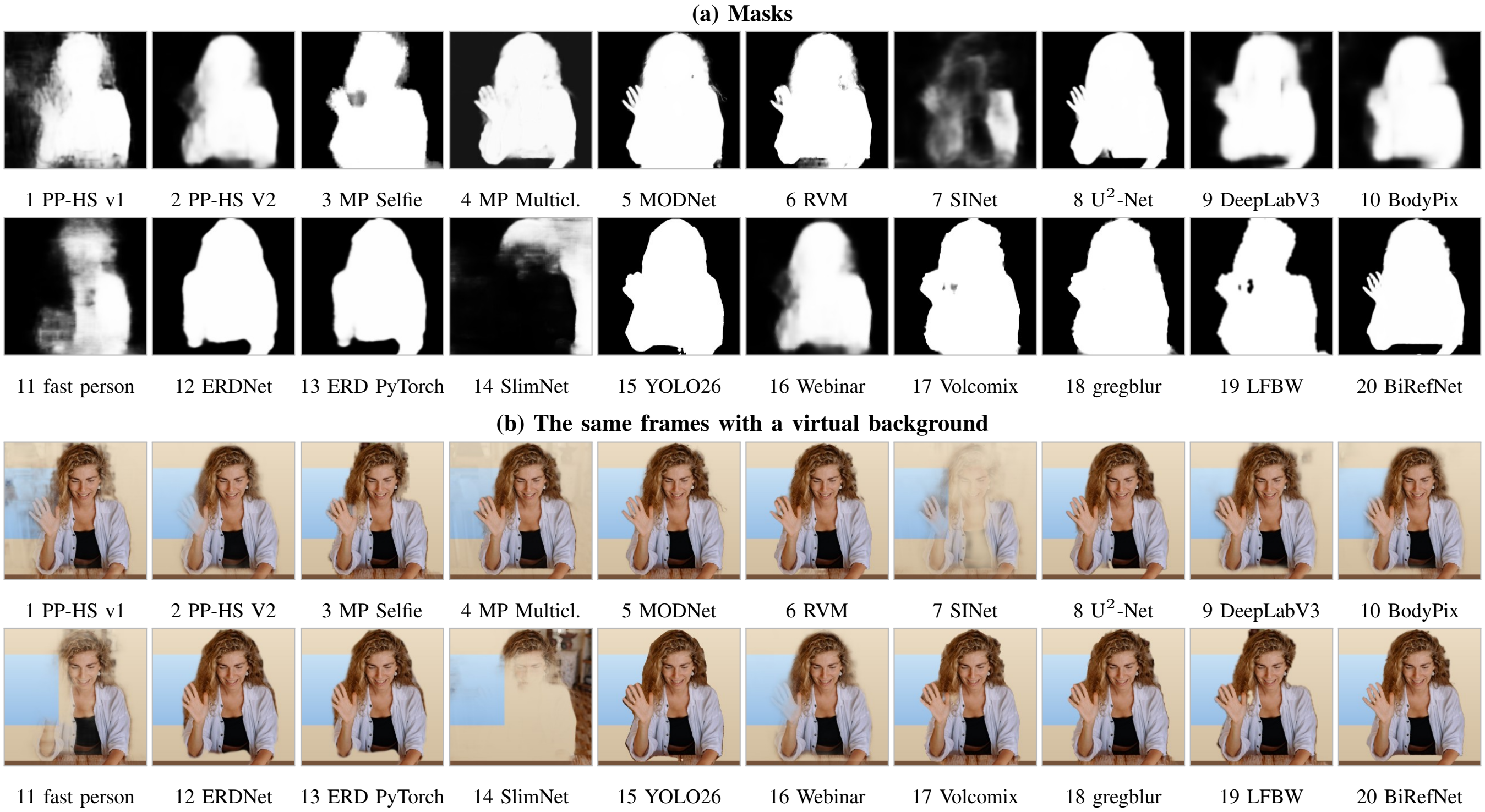

  \centering
  {\footnotesize\textbf{(a) Masks}}\\[2pt]
  \setlength{\tabcolsep}{0.8pt}
  \begin{tabular}{cccccccccc}
    \tile{0.0935\textwidth}{hand2/masks/pphsv1.jpg}{1 PP-HS v1} &
    \tile{0.0935\textwidth}{hand2/masks/pphsv2.jpg}{2 PP-HS V2} &
    \tile{0.0935\textwidth}{hand2/masks/mpselfie.jpg}{3 MP Selfie} &
    \tile{0.0935\textwidth}{hand2/masks/mpmulti.jpg}{4 MP Multicl.} &
    \tile{0.0935\textwidth}{hand2/masks/modnet512.jpg}{5 MODNet} &
    \tile{0.0935\textwidth}{hand2/masks/rvm.jpg}{6 RVM} &
    \tile{0.0935\textwidth}{hand2/masks/sinet.jpg}{7 SINet} &
    \tile{0.0935\textwidth}{hand2/masks/u2net.jpg}{8 \utwonet{}} &
    \tile{0.0935\textwidth}{hand2/masks/deeplab.jpg}{9 DeepLabV3} &
    \tile{0.0935\textwidth}{hand2/masks/bodypix.jpg}{10 BodyPix} \\[3pt]
    \tile{0.0935\textwidth}{hand2/masks/fastperson.jpg}{11 fast person} &
    \tile{0.0935\textwidth}{hand2/masks/erdnet.jpg}{12 ERDNet} &
    \tile{0.0935\textwidth}{hand2/masks/erdtorch.jpg}{13 ERD PyTorch} &
    \tile{0.0935\textwidth}{hand2/masks/slimnet.jpg}{14 SlimNet} &
    \tile{0.0935\textwidth}{hand2/masks/yolo26.jpg}{15 YOLO26} &
    \tile{0.0935\textwidth}{hand2/masks/webinar.jpg}{16 Webinar} &
    \tile{0.0935\textwidth}{hand2/masks/volc_mlkit.jpg}{17 Volcomix} &
    \tile{0.0935\textwidth}{hand2/masks/gregblur.jpg}{18 gregblur} &
    \tile{0.0935\textwidth}{hand2/masks/lfbw.jpg}{19 LFBW} &
    \tile{0.0935\textwidth}{hand2/masks/aiseg_biref.jpg}{20 BiRefNet}
  \end{tabular}
  \\[4pt]
  {\footnotesize\textbf{(b) The same frames with a virtual background}}\\[2pt]
  \setlength{\tabcolsep}{0.8pt}
  \begin{tabular}{cccccccccc}
    \tile{0.0935\textwidth}{hand2/replace/pphsv1.jpg}{1 PP-HS v1} &
    \tile{0.0935\textwidth}{hand2/replace/pphsv2.jpg}{2 PP-HS V2} &
    \tile{0.0935\textwidth}{hand2/replace/mpselfie.jpg}{3 MP Selfie} &
    \tile{0.0935\textwidth}{hand2/replace/mpmulti.jpg}{4 MP Multicl.} &
    \tile{0.0935\textwidth}{hand2/replace/modnet512.jpg}{5 MODNet} &
    \tile{0.0935\textwidth}{hand2/replace/rvm.jpg}{6 RVM} &
    \tile{0.0935\textwidth}{hand2/replace/sinet.jpg}{7 SINet} &
    \tile{0.0935\textwidth}{hand2/replace/u2net.jpg}{8 \utwonet{}} &
    \tile{0.0935\textwidth}{hand2/replace/deeplab.jpg}{9 DeepLabV3} &
    \tile{0.0935\textwidth}{hand2/replace/bodypix.jpg}{10 BodyPix} \\[3pt]
    \tile{0.0935\textwidth}{hand2/replace/fastperson.jpg}{11 fast person} &
    \tile{0.0935\textwidth}{hand2/replace/erdnet.jpg}{12 ERDNet} &
    \tile{0.0935\textwidth}{hand2/replace/erdtorch.jpg}{13 ERD PyTorch} &
    \tile{0.0935\textwidth}{hand2/replace/slimnet.jpg}{14 SlimNet} &
    \tile{0.0935\textwidth}{hand2/replace/yolo26.jpg}{15 YOLO26} &
    \tile{0.0935\textwidth}{hand2/replace/webinar.jpg}{16 Webinar} &
    \tile{0.0935\textwidth}{hand2/replace/volc_mlkit.jpg}{17 Volcomix} &
    \tile{0.0935\textwidth}{hand2/replace/gregblur.jpg}{18 gregblur} &
    \tile{0.0935\textwidth}{hand2/replace/lfbw.jpg}{19 LFBW} &
    \tile{0.0935\textwidth}{hand2/replace/aiseg_biref.jpg}{20 BiRefNet}
  \end{tabular}
  \caption{The person with the close, raised hand (\figurename~\ref{fig:handinput}c), showing her head, upper body and hand for all 20 setups, as a mask (a) and on a virtual background (b). A lost hand shows up in (b), for example as a see through ghost or as a hand cut off below the fingers. MODNet (5), RVM (6), \utwonet{} (8) and BiRefNet (20) keep separate fingers, and MediaPipe Multiclass (4) shows them on a grey background. YOLO26 (15), Volcomix (17) and LFBW (19) keep only a stub of the hand, and MediaPipe Selfie (3) a blocky part of it. Seven setups lose the hand completely, and four give a grey or noisy mask of the whole person. MODNet is shown at input 512, YOLO26 at input 640, Volcomix with the ML Kit model, gregblur with the selfie segmenter and setup 20 with BiRefNet.}
  \label{fig:handmasks2}
\end{figure*}

The portrait has no hand, so we added three free photos that look like a video call (\figurename~\ref{fig:handinput}). The first is the view the other people in a meeting usually see. A young man sits close to his front camera and waves, so his face, shoulders and open hand fill the picture, and the hand is about 320 pixels wide in a 960x1440 picture. We consider this the most important case and show it first. In the second, a man waves at his laptop from further away, and his hand is only about 130 pixels wide in a 960x1440 picture. In the third, a woman waves closer to the camera, and her open hand is raised beside her head in front of a busy room, at 1200x800. \figurename~\ref{fig:handmasks0} shows the face, shoulders and hand of the first photo in every mask, \figurename~\ref{fig:handmasks} zooms into the far hand, and \figurename~\ref{fig:handmasks2} shows the woman's head, upper body and hand.

On the front camera photo the results split into six groups.
\begin{itemize}
  \item \emph{Separate, clean fingers.} MODNet (5), RVM (6) and BiRefNet (20). MODNet at the faster input of 256 and RVM at ratio 0.4 also kept separate fingers.
  \item \emph{Fingers on a grey background.} MediaPipe Multiclass (4) shows the fingers, but its whole background is slightly grey.
  \item \emph{Fingers with some gaps filled.} Volcomix with ML Kit (17) and gregblur (18) keep the hand and the fingers, but part of the space between the fingers is filled.
  \item \emph{A blurry hand that is partly see through.} PP-HumanSeg v1 and V2 (1, 2), MediaPipe Selfie (3, blocky), SINet (7), DeepLabV3 (9), BodyPix (10) and the Webinar pipeline (16).
  \item \emph{A grey patch.} \utwonet{} (8) turns the hand into a grey square and also loses part of the forearm, although it kept separate fingers in the other two photos.
  \item \emph{The hand is gone.} fast\_person\_segmentation (11), ERDNet (12, 13), SlimNet (14), YOLO26 (15) and LFBW (19). YOLO26 and LFBW keep only a few small pieces of it.
\end{itemize}
So even a large hand close to the camera was lost or blurred by more than half of the setups.

On the far hand the results split into five groups.
\begin{itemize}
  \item \emph{Separate, clean fingers.} MODNet (5), \utwonet{} (8) and BiRefNet (20). MODNet still kept the fingers at the faster input of 256, only with softer edges.
  \item \emph{Fingers visible but soft or partly see through.} MediaPipe Multiclass (4), RVM (6), Volcomix with ML Kit (17) and gregblur (18).
  \item \emph{One blob without fingers.} PP-HumanSeg v1 and V2 (1, 2), MediaPipe Selfie (3), DeepLabV3 (9), BodyPix (10), ERDNet (12, 13), the Webinar pipeline (16) and LFBW (19).
  \item \emph{A faded, grey hand.} SINet (7), fast\_person\_segmentation (11) and SlimNet (14).
  \item \emph{Fingers cut off.} YOLO26 (15) keeps the palm but not the fingers.
\end{itemize}

The close hand was harder than the far hand, which we did not expect. Here the groups are the following.
\begin{itemize}
  \item \emph{Separate fingers.} MODNet (5), RVM (6), \utwonet{} (8) and BiRefNet (20). MediaPipe Multiclass (4) also shows the fingers, but its whole background is slightly grey.
  \item \emph{A stub of the hand without fingers.} YOLO26 (15), Volcomix with ML Kit (17) and LFBW (19). MediaPipe Selfie (3) keeps a blocky part of the hand.
  \item \emph{The hand is gone.} PP-HumanSegV2 (2), DeepLabV3 (9), BodyPix (10), ERDNet (12, 13), the Webinar pipeline (16) and gregblur (18) keep the head and body but not the raised hand.
  \item \emph{A grey or noisy mask of the whole person.} PP-HumanSeg v1 (1), SINet (7), fast\_person\_segmentation (11) and SlimNet (14).
\end{itemize}
One possible reason is that the hand is raised away from the body and in front of a busy room, so the small models take it for part of the background. We did not test this further.

Two more things stood out. BiRefNet also kept the laptop as foreground in the far photo, because it segments the main foreground objects and not only people. And ERDNet, which gave a very clean outline on the portrait, lost the fingers on all three hand photos.

\subsection{Licensing}
\label{sec:license}

The license decides whether a model can be shipped in a product. The last column of Table~\ref{tab:setups} gives the license as stated by each project. This is a summary, not legal advice, and the license of the exact files should always be checked. Four groups matter in practice.
\begin{itemize}
  \item \emph{Permissive} (Apache-2.0, MIT, BSD-3-Clause). PP-HumanSeg v1 and V2, MediaPipe Selfie and Multiclass (see their model cards), MODNet, SINet, \utwonet{}, DeepLabV3, BodyPix, SlimNet, gregblur, the Volcomix code with the ML Kit model, and BiRefNet.
  \item \emph{GPL-3.0}. RVM and the Linux-Fake-Background-Webcam project.
  \item \emph{AGPL-3.0}. Ultralytics YOLO. A paid license is offered for closed source use.
  \item \emph{Unclear}. fast\_person\_segmentation (its repository has no license, the original model is MIT), ERDNet and ERD seg-matting (the model's origin has no license), the Google Meet models in Volcomix (first shared as Apache-2.0, later under Google terms), and the webinar repository (no license, while its model is Apache-2.0).
\end{itemize}
The AI-Segmenter app itself is MIT, but its default pipeline uses RVM and YOLO.
\section{Practical Observations}
\label{sec:findings}

While running the setups we found several things that mattered a lot, apart from the choice of model.

\subsection{The runtime library can change speed many times over}
The same PP-HumanSegV2 network took about 33~ms per frame in PaddlePaddle and about 4.4~ms in ONNX Runtime, measured in setup 16 without its temporal and edge steps. The Mobile-UNet of setup 11 took about 250~ms per frame in OpenCV 5.0 and 11 to 17~ms in OpenCV 4.11, in a separate test that compared the two versions. Before calling a model slow, it is worth trying another runtime or another version.

\subsection{The steps around the model cost time}
The Webinar HumanSeg pipeline runs at 86--89~FPS with the raw model mask and at 33--35~FPS with its temporal and edge steps. Most of the extra time goes to the edge step, which looks at every pixel near the person's edge. The same pattern holds for the Volcomix and gregblur pipelines. Each step of a pipeline should be measured on its own.

\subsection{Per pixel steps need compiled code in Python}
Edge filters that run on the GPU in a browser are slow when written with plain NumPy on a CPU. The joint bilateral upsampling of the webinar pipeline took 150 to 300~ms per frame in NumPy and about 14~ms as a compiled Numba~\cite{numba} loop with 4 threads.

\subsection{A bigger input does not help a fixed size model}
We fed PP-HumanSegV2, which was trained at 256x144, a larger picture of 512x288 and 640x360. The mask got worse, not better. More of the flag leaked in and a hole appeared in the body. For more detail, a model made for a larger input is the better choice.

\subsection{Small models lose thin parts}
At 256x144 the finger of a hand further from the camera is only a few model pixels wide and can vanish, and thin hair has the same problem. We first saw this in live use of the Webinar HumanSeg pipeline. The three hand photos confirmed it (Section~\ref{sec:hands}). Separate fingers came only from MODNet, RVM, \utwonet{} and BiRefNet, while most small segmentation models blurred the hand, turned it into a blob or lost it. This happened even with the large hand of the front camera photo, so the input size is not the only reason.

\subsection{Check what the output means}
Two of the pipelines did not do what their settings suggest in our tests. In gregblur, the selfie segmenter option treats the model's person mask as the background, so the person would be blurred. In Linux-Fake-Background-Webcam, the default smoothing over time has no effect, because the old mask and the new mask use the same memory. In our runs we used the selfie segmenter mask the correct way and left the smoothing off. Small mistakes like these are easy to miss, so the mask itself should always be checked and not only the final picture.

\section{Model Selection Guidelines}
\label{sec:guide}

Table~\ref{tab:guide} gives a short answer for common needs. The paragraphs below explain each answer. All speeds are for our laptop CPU at 640x480, so a weaker CPU needs some extra room.

\begin{table*}[!t]
  \centering
  \caption{Selection guide. FPS is for the whole pipeline on an Intel Core i7-13620H laptop CPU at 640x480 (test clip, then webcam).}
  \label{tab:guide}
  \footnotesize
  \renewcommand{\arraystretch}{1.2}
  \setlength{\tabcolsep}{4pt}
  \begin{tabularx}{\textwidth}{@{}p{4.4cm} p{4.6cm} p{1.9cm} Y@{}}
    \toprule
    \textbf{Need} & \textbf{Recommended setup} & \textbf{FPS} & \textbf{Watch out for} \\
    \midrule
    Weak or old computer & MediaPipe Selfie, landscape (3) & 110 / 41--63 & Busy backgrounds can leak in as solid patches, blocky hands \\
    Good quality and fast & ncnn ERDNet (12) & 47 / 38--43 & License unclear, loses fingers and raised hands \\
    Good quality, fast, permissive license & Volcomix with ML Kit selfie (17) & 45--47 / 42 & More steps to implement, fingers partly joined or lost \\
    Steady mask, edge that follows the hair & Webinar HumanSeg pipeline (16) & 33--35 / 34 & Its code has no license, raised hands were blurred, a blob or lost \\
    Best soft edges in real time & RVM, ratio 0.4 (6) & 25--29 & GPL-3.0 \\
    Soft edges and fingers, permissive license & MODNet, input 256 (5) & 20 & Less detail than at input 512 \\
    People behind the speaker & YOLO26n-seg at 320, biggest person only (15) & 23 / 32 & AGPL-3.0, hard edges that cut the hair \\
    Photos, not live video & BiRefNet (20) or \utwonet{} (8) & 0.1 and 1.9 & Far too slow for a call on a CPU \\
    \bottomrule
  \end{tabularx}
\end{table*}

\subsection{Weak or old computers}
MediaPipe Selfie (landscape) was the fastest setup by a wide margin, at 110~FPS on the test clip and 41 to 63~FPS on the webcam, and its model is only 0.25~MB. The outline is sharp, but a busy background can leak in as a solid patch (\figurename~\ref{fig:flag}). The Linux-Fake-Background-Webcam recipe uses the same model and is also fast, but its hard threshold keeps such a patch fully solid, so the plain model or a softer clean up is the better choice.

\subsection{Good quality in real time on a normal laptop}
ncnn ERDNet gave a clean mask on the portrait, close to MODNet, at 47~FPS on the test clip and about 40 on the webcam. But it lost the raised hand completely in the front camera and close photos, turned the far hand into a blob, kept one stripe of the flag next to the arm, and the license of the model is unclear. With a clear license, the Volcomix pipeline with the ML Kit selfie model was the strongest fast option in our test. It reached 45 to 47~FPS on the test clip and 42 on the webcam, and its mask was clean on the test image apart from a small gap near the arm. On the front camera photo it kept the fingers, with some gaps between them filled, and on the far hand it showed the fingers softly. Of the close raised hand it kept only a stub without fingers. Plain PP-HumanSegV2 is another permissive choice at 26 to 27~FPS, with a faint leak of busy backgrounds, but it did not keep raised hands well.

\subsection{Hair, fingers and soft edges}
MODNet and RVM keep loose hair strands (\figurename~\ref{fig:hair}). In our hand photos MODNet kept separate fingers all three times, and RVM in the front camera and close photos (Figs.~\ref{fig:handmasks0}, \ref{fig:handmasks} and~\ref{fig:handmasks2}). RVM at ratio 0.4 runs at 25 to 29~FPS, has soft edges like MODNet and stays steady on video because it remembers earlier frames, but it is GPL-3.0. MODNet is Apache-2.0 and runs at about 20~FPS at input 256, where it still kept separate fingers on the front camera and far hand photos. At its default input of 512 it looks better but reaches only about 6~FPS. Among the setups with a permissive license that run in or close to real time, MODNet at 256 was the only one that kept separate fingers on the far hand photo, so it is our suggestion when hands and fingers matter.

\subsection{A flickering mask or rough edges}
Before changing the model, it is worth adding the steps of \figurename~\ref{fig:pipeline}. In the Webinar HumanSeg pipeline, a motion based temporal step and joint bilateral upsampling make the PP-HumanSegV2 mask flicker less and follow the hair more closely, and it still runs at 33 to 35~FPS. These steps did not help with raised hands, which the pipeline blurred, turned into a blob or lost in our photos. Its code has no license, so a product needs its own version of these well known steps.

\subsection{Closed source products}
The permissive group of Section~\ref{sec:license} is the safe choice. Real time options in that group are MediaPipe Selfie, PP-HumanSeg v1 and V2, SlimNet, the Volcomix pipeline with ML Kit, gregblur with the selfie segmenter and SINet. SlimNet drops to about 16~FPS and gregblur to about 18 to 19~FPS on the webcam. MODNet at input 256 is close to real time. RVM (GPL-3.0) and YOLO (AGPL-3.0) need a commercial license where one is offered, and models with an unclear license are best avoided.

\subsection{People behind the speaker}
The plain segmentation and matting models treat everyone in the frame as one person. YOLO26n-seg finds each person on their own, so it can keep only the biggest one. At input 320 it ran at 23~FPS on the test clip and 32 on the webcam. Its edges are hard and cut the hair, and it is AGPL-3.0. Another way is to run a separate person or face detector next to a segmentation model, but we did not measure such a setup in this study.

\subsection{Blur only}
In our experience blur hides small mask errors better than a new image, because the blurred room still looks like the room. A fast model with a slightly rough edge, such as MediaPipe Selfie, is often enough. The gregblur pipeline is designed for blur and avoids a halo around the person, at about 23~FPS on the test clip with the selfie segmenter.

\subsection{Photos instead of live video}
For single images the slow models are fine. BiRefNet gave the cleanest edge on the test image but took 9 to 23 seconds per frame on the CPU. \utwonet{} took about half a second and gave a clean, smooth outline. MediaPipe Multiclass took about 0.13 seconds and removed most of the flag, with a faint grey background value.

\subsection{Setups to avoid}
BodyPix gave the weakest mask, with a see through leak and grey patches inside the body, and its authors have marked it as deprecated. DeepLabV3 is a general model and gave a soft, wide edge even though it does not leak the background. \utwonet{} and BiRefNet are too slow for a call on a CPU.

\section{Limitations}
\label{sec:limits}

\begin{itemize}
  \item \emph{One machine.} All numbers come from one Windows laptop with an Intel Core i7-13620H. Other CPUs give other numbers. We expect the order of the setups to stay roughly the same, but we did not check this.
  \item \emph{CPU only.} With a GPU, slow models such as MODNet at 512, \utwonet{} and BiRefNet would likely be much faster, but we did not measure this.
  \item \emph{No accuracy score.} We had no ground truth masks, so quality was judged visually on four photos, one test clip made from the portrait, and live webcam use. A dataset based score, for example on the teleconferencing dataset of PP-HumanSeg~\cite{chu2021pphumanseg}, would make the quality ranking stronger.
  \item \emph{Few scenes.} Our inputs show one portrait with a flag, three video call photos with a raised hand, and one person in an office. Low light, strong back light, long hair, fast motion and several people were not tested in a controlled way.
  \item \emph{Our own versions of the pipelines.} Setups 16 to 20 are Python versions of projects that run in a browser, on Linux or on an NVIDIA GPU. They follow the published steps and settings, but they are not the original programs, and their speed is not the speed of the original apps.
  \item \emph{Run to run variation.} The same setup could vary by up to two times between runs, so small differences between setups should not be over interpreted.
  \item \emph{Other systems.} We did not test on Linux or macOS directly.
\end{itemize}

\section{Conclusion}
\label{sec:conclusion}

We ran 20 open person segmentation setups for video call backgrounds on the same laptop CPU and compared their speed, their masks and their licenses. Real time on a CPU is easy to reach with small models such as MediaPipe Selfie, ncnn ERDNet, the ML Kit model in the Volcomix pipeline and PP-HumanSegV2. The softest edges and the best hair came from matting models such as RVM and MODNet, which reach or come close to real time only at a smaller input. Raised hands and fingers are the hardest part. On our three video call photos only MODNet and BiRefNet kept separate fingers every time. RVM did on the front camera and close photos, and \utwonet{} on the far and close photos. Most fast segmentation models blurred the hand, turned it into a blob or lost it, even in the close front camera view. Only YOLO in our test gives a separate mask for each person, which helps when others are in the room, but its license and hard edges limit its use. The runtime library and the steps around the model can also change the speed by several times.

We hope the selection guide in Section~\ref{sec:guide} helps developers pick a model without running all of them first. In future work we would like to add a dataset based accuracy score, more scenes, GPU and mobile runs, and the options we could not test here.

\section*{Image Credits}
The test image is the NASA portrait of Kalpana Chawla in an orange suit, a public domain photograph, taken from Wikimedia Commons. The front camera photo is ``Man with Hand Raised'' by ANTONI SHKRABA production, used under the Pexels License (\url{https://www.pexels.com/photo/man-with-hand-raised-8374247/}). The far hand photo is ``A Man Using a Laptop in Video Calling'' by Vlada Karpovich, used under the Pexels License (\url{https://www.pexels.com/photo/a-man-using-a-laptop-in-video-calling-4939622/}). The close hand photo is ``Photo of Woman Saying Hi Through Laptop'' by Yan Krukau, used under the Pexels License (\url{https://www.pexels.com/photo/photo-of-woman-saying-hi-through-laptop-4458423/}).

\bibliographystyle{IEEEtran}
\bibliography{references}

\end{document}